\documentclass[groupaddress
 aip,
 amsmath,amssymb,
 preprint,%
]{revtex4-1}

\usepackage{graphicx}
\usepackage{dcolumn}
\usepackage{bm}
\usepackage{float}
\usepackage{amsmath, amssymb, amsthm, latexsym}
\usepackage{xcolor}
\usepackage[utf8]{inputenc}
\usepackage[T1]{fontenc}
\usepackage{mathptmx}
\usepackage{etoolbox}
\usepackage{array}
\usepackage{comment}
\usepackage{enumitem}
\usepackage{subcaption}
\usepackage[hidelinks]{hyperref}
\usepackage{booktabs}
\usepackage{tabularx}

\makeatletter
\def\@email#1#2{%
 \endgroup
 \patchcmd{\titleblock@produce}
  {\frontmatter@RRAPformat}
  {\frontmatter@RRAPformat{\produce@RRAP{*#1\href{mailto:#2}{#2}}}\frontmatter@RRAPformat}
  {}{}
}%
\makeatother
\begin{document}


\title{Measurement of Kerr rotation using a variable-angle polarizer method
}
\author{Shusmita Podder Pooza}
\author{Hailey Cossey}
\author{Dipanjan Mazumdar\textsuperscript{*}}
\email[Corresponding author: ]{dmazumdar@siu.edu}
\affiliation{ 
School of Physics and Applied Physics, Southern Illinois University, Carbondale, IL, 62901 
}

\date{\today}

\begin{abstract}
The magneto-optical Kerr effect (MOKE) occurs when polarized light reflects from a magnetized surface, causing a small change in the polarization angle and state. Measurement of the rotation in the angle (Kerr rotation) is well established and typically performed close to the null configuration in a polarizer-analyzer geometry. However, accurate measurement always remains a challenge and as the effect depends intricately on several optical parameters. Here we performed a series of longitudinal magneto-optical Kerr effect (MOKE) measurements on $p$ and $s$ polarized laser light at various polarizer angles on a Cobalt thin film and measured the orthogonal components of the reflected polarized light using a Wollaston prism. Analytical expressions for the orthogonal light components were fitted to the average intensity and the MOKE signal measured at different polarizer angles to obtain the Kerr roations. Our analysis yielded a $p (s)$-Kerr rotation of 0.46 (0.65) milliradians for a 633~nm laser at 45$^\circ$ angle of incidence, which agrees very well with our estimated value for Co using available literature data. Apart from being very accurate, the advantage of the process is that it eliminates the inherent uncertainties in single-point measurements of the Kerr rotation. 
\end{abstract}

\maketitle

\section{Introduction}

 Magneto-optical Kerr effect (MOKE) has emerged as a popular and effective tool for studying magnetism and magnetic material properties\cite{freiser1968survey, qiu1999surface} and provides a sensitive and non-destructive method for investigating magnetic anisotropy, \cite{lehnert2010magnetic, he2015probing, wohlrath2025quadratic} domain behavior, \cite{mccord2015progress,jiang2013direct, kim2019micromagnetometry,reichlova2019imaging} in ferromagnetic materials and successfully extended to antiferromagnetic materials. \cite{nvemec2018antiferromagnetic,farhang2026topological, okamura_giant_2026,sunko_magneto-optical_2026,higo_large_2018} It is, therefore, not surprising that MOKE is now an established microscopy \cite{kim2020extreme,huang2026advancing} and imaging \cite{sung2018magnetic} tool, and is being developed for several other applications \cite{haider2017review}.

 Kerr rotation, polarization rotation of reflected light, is one of the most important quantities of interest in a MOKE measurement and is typically of the order of milliradians or less in magnetic materials. While MOKE is straightforward to observe, accurate measurement of Kerr rotation is a non-trivial proposition, and researchers continue to explore new methods \cite{suzuki2024measurement} or enhance the Kerr signal using optical methods such as anti-reflection, \cite{kim2020extreme}  and interference. \cite{sumi_interference_2018}

\begin{figure}[htbp]
    \centering
    \includegraphics[width=0.7\linewidth]{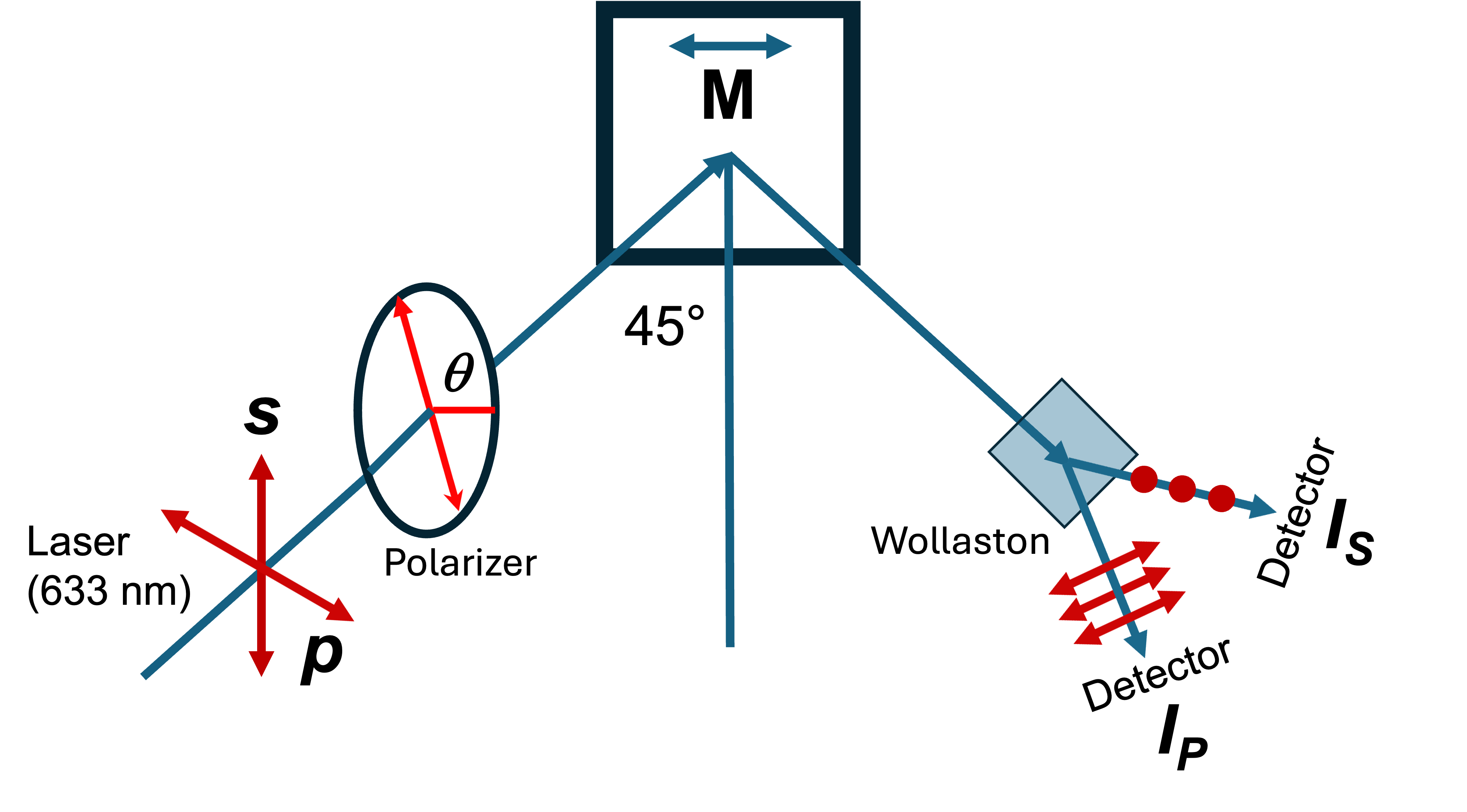}
    \caption{Schematic diagram of the longitudinal (MOKE) setup employed in this work. A 633~nm $s$ and $p$-polarized laser beam passes through variable-angle polarizer and is reflected from the sample. The $s$ and $p$ components of the reflected light are detected using a Wollaston prism}
    \label{fig:MOKE_CONFI}
\end{figure}

In this work, we demonstrate an accurate method for estimating the Kerr rotation of thin film materials utilizing a variable-angle polarizer and a Wollaston prism in a longitudinal configuration, as shown in Figure \ref{fig:MOKE_CONFI}. By meausring annd analyzing the magnetic hysteresis measurements as a function of the polarizer angle, $\theta$, for $ p$ and $ s$-polarized laser, and fitted the orthogonal intensities and MOKE signals to equations obtained from Jones matrix analysis we were abale to deduce accurately the $p$ and $s$ Kerr rotation of the investigated Cobalt film. Using different laser and detector configurations, we obtain the Kerr rotations from the fit parameters, leading to a highly accurate method. The advantage of the multi-angle fitting process is that it does not rely on a single point such as the null-point for the Kerr rotation estimation and alleviates the requirement of precise optical alignment.  Using this method, we obtained a p-Kerr rotation of 0.46 milliradian and an $s$-Kerr rotation of 0.65 milliradian for a thick amorphous Cobalt film, which is in excellent agreement with expected values. 


\section{Methodology}

The magneto-optical Kerr effect (MOKE) was measured in the longitudinal configuration shown in Fig. \ref{fig:MOKE_CONFI} using a 0.9~mW laser diode as the light source. Laser light is passed through a 100000:1 Glan Thompson polarizer and then a focusing lens before being incident at a 45-degree angle on a sample + electromagnet assembly. The electromagnets (Kroncke 5A) are powered by a 400 W (75V/6A) Kepco Bipolar Operational Power supply. A mild vacuum is applied to hold the sample against the stage through a small hole at the center of the stage. The beam reflects off the sample, passes through a chopper and a Wollaston prism, and is then collected by a photodetector and a lock-in amplifier interfaced with a data acquisition device via LabVIEW.

\section{Jones Matrix analysis of Longitudinal MOKE setup}

The experimental setup used in this work, as in Fig. \ref{fig:MOKE_CONFI}, can be modeled as: \par
\vspace{0.5cm}
$\textbf{Laser} \rightarrow \textbf{Polarizer} \rightarrow \textbf{Sample} \rightarrow \textbf{Beam Splitter}+\textbf{Photodetector}$ \par
\vspace{0.5cm}

Here, we represent each optical element with its corresponding Jones matrix and
\begin{equation}
\mathbf{E}_f = SP \, \mathbf{E}_i
\label{setup_writing}
\end{equation}
The electric field vector amplitudes, $E_i$ and $E_f$ initial and final, represent the incident and reflected/final electric fields, while $S$ and $P$ are the sample and the polarizer matrices.
By performing straightforward matrix multiplication, we can obtain the expressions for the $p$ and $s$ components of the final electric field $(E_{fp})$ and $(E_{fs})$ respectively.

\begin{align}
\mathbf{E}_{\text{f}} &=
\begin{pmatrix}
E_{fp} \\
E_{fs}
\end{pmatrix}=
\begin{pmatrix}
r_p e^{i\delta_p} & r_{ps} e^{i\delta_{ps}} \\
r_{sp} e^{i\delta_{sp}} & r_s e^{i\delta_s}
\end{pmatrix} \nonumber \\
&\quad \times
\begin{pmatrix}
\cos^2\theta & \cos\theta \sin\theta \\
\cos\theta \sin\theta & \sin^2\theta
\end{pmatrix}
\begin{pmatrix}
\cos\alpha \\
\sin\alpha
\end{pmatrix}
\label{E_final_matrix}
\end{align}

where $\alpha$ indicates the laser polarization, which is set either to the $p$($\alpha =0$) or $s$ ($\alpha = \pi/2$) mode. With straightforward matrix multiplication, we have:

\begin{align}
E_{fp} &= \left( r_p e^{i\delta_p} \cos^2\theta + r_{ps} e^{i\delta_{ps}} \cos\theta \sin\theta \right)\cos\alpha \nonumber \\
&\quad + \left( r_p e^{i\delta_p} \cos\theta \sin\theta + r_{ps} e^{i\delta_{ps}} \sin^2\theta \right)\sin\alpha
\label{E_fH_ini}
\end{align}

\begin{align}
E_{fs} &= \left( r_{sp} e^{i\delta_{sp}} \cos^2\theta + r_s e^{i\delta_s} \cos\theta \sin\theta \right)\cos\alpha \nonumber \\
&\quad + \left( r_{sp} e^{i\delta_{sp}} \cos\theta \sin\theta + r_s e^{i\delta_s} \sin^2\theta \right)\sin\alpha
\label{E_fV_ini}
\end{align}

In our experiments, we detect the orthogonal $s$ and $p$ intensities. 
 \begin{align*}
 I_p = |E_{fp}|^2  \quad \text{and} \quad  I_s = |E_{fs}|^2 
\end{align*}

\subsection{Orthogonal intensity calculations for $p$-polarized laser}
For $\alpha = 0^\circ$, eqns. \ref{E_fH_ini} and \ref{E_fV_ini}
\begin{align}
E_{fp} &= r_p e^{i\delta_p} \cos^2\theta + r_{ps} e^{i\delta_{ps}} \cos\theta\sin\theta 
\label{E_FH_0}
\\
E_{fs} &= r_s e^{i\delta_s} \cos\theta\sin\theta + r_{sp} e^{i\delta_{sp}} \cos^2\theta
\label{E_FV_0}
\end{align}
Now, taking the conjugate of equation \ref{E_FH_0} and \ref{E_FV_0}
\begin{align}
E_{fp} E_{fp}^* &=
\left( r_p e^{i\delta_p} \cos^2\theta + r_{ps} e^{i\delta_{ps}} \cos\theta\sin\theta  \right) \nonumber \\
&\quad \times
\left( r_p e^{-i\delta_p} \cos^2\theta+ r_{ps} e^{-i\delta_{ps}} \cos\theta\sin\theta  \right)
\end{align}
\begin{align}
E_{fs} E_{fs}^* &=
\left( r_{sp} e^{i\delta_{sp}} \cos\theta \sin\theta + r_s e^{i\delta_s} \sin^2\theta \right) \nonumber \\
&\quad \times
\left( r_{sp} e^{-i\delta_{sp}} \cos\theta \sin\theta + r_s e^{-i\delta_s} \sin^2\theta \right)
\end{align}
\[
 \text{where } \quad \Delta= \delta_{sp}- \delta_{s} \quad \text{or} \quad \Delta= \delta_{p}- \delta_{ps}
 \] 
Multiply these terms and get,

 \begin{align}
I_{pp} = |E_{fp}|^2 &= \frac{1}{4} r_{ps}^2 \sin^2(2\theta) + r_p^2 \cos^4\theta \nonumber \\
&\quad + r_p r_{ps} \cos^2\theta \sin(2\theta)\cos\Delta
\label{final I_H_0}
\end{align}
\begin{align}
I_{sp} = |E_{fs}|^2 &= \frac{1}{4}r_{s}^2 \sin^22\theta + r_{sp}^2 \cos^4\theta \nonumber \\
&\quad + r_{sp} r_s \sin(2\theta) \cos^2\theta \cos\Delta
\label{final_IV_0}
\end{align}

Defining, the $p$-Kerr rotation as,
\begin{equation}
\theta_k^p=\frac{r_{ps}}{r_p}\cos\Delta,
\end{equation}

\begin{equation}
    \boxed{I_{pp}=r_{p}^2 \cos^4\theta + \frac{1}{4}r_{ps}^2 \sin^2(2\theta) +  r_p^2 \theta_k^p \sin(2\theta) \cos^2\theta}
\end{equation}

Similarly,
   \begin{equation}
    \boxed{I_{sp}=\frac{1}{4} r_{s}^2 \sin^2(2\theta) + r_{sp}^2 \cos^4\theta+ r_s^2 \theta_k^s \sin(2\theta) \cos^2\theta }
   \label{final E_fv 0}
\end{equation}
\text{where } ${\theta_k^s} = \frac{r_{sp}}{r_s} \cos\Delta$ 
is the $s$-Kerr rotation.

Note that, in our notation, $I_{sp}$ denotes the $s$-intensity measured at the detector for a $p$-polarized laser.

\subsection{Orthogonal MOKE signals for $p$-polarized laser}
Only the Kerr angle, $\theta_K$, changes sign under a reversal of magnetization, \textbf{M}, resulting in the magneto-optical Kerr effect (MOKE) signal, $\Delta I_{pp}$ or  $\Delta I_{sp}$. For example,
\begin{equation}
\Delta I_{sp} = I_{sp}(+M) - I_{sp}(-M) 
\label{DelI_v}
\end{equation}

The magneto-optical Kerr effect (MOKE) signal, $\Delta I_{sp}$,  from equation \ref{final_IV_0} is given by:
\begin{equation}
    \Delta I_{sp} = 2r_s^2 \left( \frac{r_{sp}}{r_s} \cos\Delta \right) \sin2\theta \cos^2\theta
\end{equation}
\text{where } ${\theta_k^s} = \frac{r_{sp}}{r_s} \cos\Delta$
\begin{equation}
    \boxed{\Delta I_{sp} = 2\theta_k^s r_s^2 \sin2\theta \cos^2\theta}
    \label{eqn:del_Isp}
\end{equation}
Now, the MOKE signal for \( \Delta I_{pp} \) from equation \ref{final I_H_0},
\begin{equation}
    \Delta I_{pp} = r_p^2 \left( \frac{r_{ps}}{r_p} \cos \Delta \right ) \sin 2\theta \cos^2 \theta
    \label{DelI_H initial for 0}
\end{equation}
\begin{equation}
   \boxed{\Delta I_{pp} = 2{r_p^2}\theta_k^p \sin 2\theta \cos^2 \theta} 
    \label{Del I_H final for 0}
\end{equation}
\text{where } ${\theta_k^p} = \frac{r_{ps}}{r_p} \cos\Delta$
\subsection{Orthogonal intensity calculations for $s$-polarized laser}

when $\alpha = 90^\circ$, $(E_{fp})$ and $(E_{fs})$  becomes (eqns \ref{E_fH_ini} and \ref{E_fV_ini})
\begin{align}
E_{fs} = r_{sp} e^{i\delta_{sp}} \cos\theta \sin\theta + r_s e^{i\delta_s} \sin^2\theta \quad
\label{E_fV 90deg}\\
E_{fp} =r_p e^{i\delta_p} \cos\theta \sin\theta + r_{ps} e^{i\delta_{ps}} \sin^2\theta
\label{E_fH 90 degree}
\end{align}
taking the conjugate of eqn \ref{E_fV 90deg},
\begin{align}
E_{fs} E_{fs}^* &=
\left( r_{sp} e^{i\delta_{sp}} \cos\theta \sin\theta + r_s e^{i\delta_s} \sin^2\theta \right) \nonumber \\
&\quad \times
\left( r_{sp} e^{-i\delta_{sp}} \cos\theta \sin\theta + r_s e^{-i\delta_s} \sin^2\theta \right)
\label{E_fv_initial_conjugate_term}
\end{align}
Simplifying, eqn. \ref{E_fv_initial_conjugate_term} becomes, 
\begin{align}
    E_{fs} E_{fs}^* = r_{sp}^2 (\cos\theta \sin\theta)^2 
+ r_s^2 (\sin^2\theta)^2 \\
+ 2 r_{sp} r_s \cos\theta \sin^3\theta \cos(\delta_{sp} - \delta_s)\quad
\label{E_fv final term}
\end{align}
Therefore, 
\begin{equation}
    \boxed{I_{ss} = \frac{1}{4} r_{sp}^2 \sin^2(2\theta) + r_{s}^2 \sin^4\theta + r_{s}^2 \theta_k^s \sin(2\theta) \sin^2\theta
    \label{I_v final}}
\end{equation}
Similarly,
\begin{align}
E_{fp} E_{fp}^*
&=
\left( r_p e^{i\delta_p} \cos \theta \sin \theta + r_{ps} e^{i\delta_{ps}} \sin^2 \theta \right) \nonumber \\
&\quad \times
\left( r_p e^{-i\delta_p} \cos \theta \sin \theta + r_{ps} e^{-i\delta_{ps}} \sin^2 \theta \right)
\label{E_fH_initial}
\end{align}
\begin{equation}
    \boxed{I_{ps} = \frac{1}{4} r_{p}^2 \sin^2(2\theta) + r_{ps}^2 \sin^4\theta + r_p \theta_k^p \sin(2\theta) \sin^2\theta}
    \label{I_ps-final}
\end{equation}

\subsection{Orthogonal MOKE signals for $s$-polarized laser}

From equation \ref{I_v final},

\begin{equation}
   \boxed{\Delta I_{ss} =  2{r_s^2}\theta_k^s \sin 2\theta \sin^2 \theta} 
    \label{I_V final eq 2}
\end{equation}
Now, the MOKE signal for \( \Delta I_{ps} \) from equation \ref{I_ps-final},
\begin{equation}
   \boxed{ \Delta I_{ps} = 2r_p^2 \theta_k^p \sin 2\theta \sin^2 \theta}
    \label{DelI_H initial}
\end{equation}

\begin{table}[H]
\centering
\renewcommand{\arraystretch}{1.4}
\setlength{\tabcolsep}{6pt}

\begin{tabular}{|c|c|c|}
\hline
 & \textbf{\textit{$\bm{p}$-laser}} & \textbf{\textit{$\bm{s}$-laser}} \\
\hline

\textbf{\textit{$\bm{p}$-intensity}} &
\parbox{4cm}{
\[
r_p^2\cos^4\theta
+\frac{1}{4}r_{ps}^2\sin^2(2\theta)
\]
\[
+r_p^2\theta_k^p\sin(2\theta)\cos^2\theta
\]
}
&
\parbox{4cm}{
\[
\frac{1}{4}r_p^2\sin^2(2\theta)
+r_{ps}^2\sin^4\theta
\]
\[
+r_p^2\theta_k^p\sin(2\theta)\sin^2\theta
\]
}
\\
\hline

\textbf{\textit{$\bm{p}$-MOKE}} &
\parbox{4cm}{
\[
2r_p^2\theta_k^p
\sin(2\theta)\cos^2\theta
\]
}
&
\parbox{4cm}{
\[
2r_p^2\theta_k^s
\sin(2\theta)\sin^2\theta
\]
}
\\
\hline

\textbf{\textit{$\bm{s}$-intensity}} &
\parbox{5cm}{
\[
\frac{1}{4}r_s^2\sin^2(2\theta)
+r_{sp}^2\cos^4\theta
\]
\[
+r_s^2\theta_k^s
\sin(2\theta)\cos^2\theta
\]
}
&
\parbox{5cm}{
\[
r_s^2\sin^4\theta
+\frac{1}{4}r_{sp}^2\sin^2(2\theta)
\]
\[
+r_s^2\theta_k^s
\sin(2\theta)\sin^2\theta
\]
}
\\
\hline

\textbf{\textit{$\bm{s}$-MOKE}} &
\parbox{5cm}{
\[
2r_s^2\theta_k^s
\sin(2\theta)\cos^2\theta
\]
}
&
\parbox{5cm}{
\[
2r_s^2\theta_k^s
\sin(2\theta)\sin^2\theta
\]
}
\\
\hline

\end{tabular}

\caption{Angular dependence of intensity and MOKE signals as a function of the polarizer angle, $\theta$, as shown in Fig. \ref{fig:MOKE_CONFI}. An overall normalization factor is omitted here.}
\label{Table:alpha_angles}

\end{table}

\section{Estimation of $p$ and $s$ Kerr rotation}
You and Shin \cite{you1998generalized} provided simplified expressions for the complex Kerr angle. For the longitudinal configuration where $m_y =1$ and $m_x=m_z=0$, assuming a thick magnetic media, we have, for $p$-polarized incident light

\begin{equation}
\tilde{\theta}_k^{p} \equiv 
\left( \frac{r_{sp}}{r_{pp}} \right)^{\text{long.}}
= \frac{\cos \theta_{0}\, \tan \theta_{1}}{\cos(\theta_{0} + \theta_{1})}
\cdot \frac{i\, n_{0} n_{1} Q}{\left( n_{1}^{2} - n_{0}^{2} \right)}
\label{p-Kerr-rotation-You-Shin}
\end{equation}
Similarly, for $s$-polarized light,
\begin{equation}
\tilde{\theta}_k^{s} \equiv 
\left( \frac{r_{sp}}{r_{pp}} \right)^{\text{long.}}
= \frac{\cos \theta_{0}\, \tan \theta_{1}}{\cos(\theta_{0} - \theta_{1})}
\cdot \frac{i\, n_{0} n_{1} Q}{\left( n_{1}^{2} - n_{0}^{2} \right)}
\label{s-Kerr-rotation-You-Shin}
\end{equation}

The denominator
$(n_{1}^{2}-n_{0}^{2})$ describes the optical contrast at the interface and
determines the sensitivity of the Kerr rotation and ellipticity. The prefactor
$\cos \theta_{0}\tan \theta_{1} / \cos(\theta_{0}+\theta_{1})$ represents the geometrical dependence of the Kerr effect on the angle of incidence. Thus, the
measured Kerr signal depends on both the intrinsic magneto-optical properties
of the material and the optical geometry of the experiment.Knowing Q and the refractive indices of the magnetic medium, we can determine the expected Kerr rotation and ellipticity of a material.\cite{carey_magneto-optic_1978} For cobalt at a wavelength of 633\,nm, the complex refractive index n, and magneto-optical constant Q are taken as $n_0 = 1$,
$n_1 = 2.214 + i\,4.174$, \cite{RefractiveIndexINFO2025}
Q = 0.0275 - i\,0.006,\cite{carey_magneto-optic_1978}
$\theta_0 = 45^\circ$

\begin{table}[ht]
\centering
\renewcommand{\arraystretch}{1.6}      
\setlength{\tabcolsep}{12pt}           
\begin{tabular}{ccc}
\hline
Laser & Complex Kerr angle & Kerr Rotation \\
\hline
$p$ & $\tilde{\theta}_k^{p}=0.4886-i0.7703~\text{mrad}$ & $\theta_k^{p}=0.4886~\text{mrad},\ \epsilon_k^{p}=-0.7703~\text{mrad}$ \\
$s$ & $\tilde{\theta}_k^{s}=0.5877-i0.5385~\text{mrad}$ & $\theta_k^{s}=0.5877~\text{mrad},\ \epsilon_k^{s}=-0.5385~\text{mrad}$ \\
\hline
\end{tabular}
\caption{Kerr rotation and ellipticity estimates for $s$- and $p$-polarized laser on Cobalt.}
\label{tab:kerr_estimates}
\end{table}

 We shall next explain our Kerr rotation measurements for $p$ and $s$-polarized laser and compare with our estimates as shown here.

\section{Variable-angle polarizer measurements}

\begin{figure}[h]
    \centering
\includegraphics[width=0.8\linewidth]{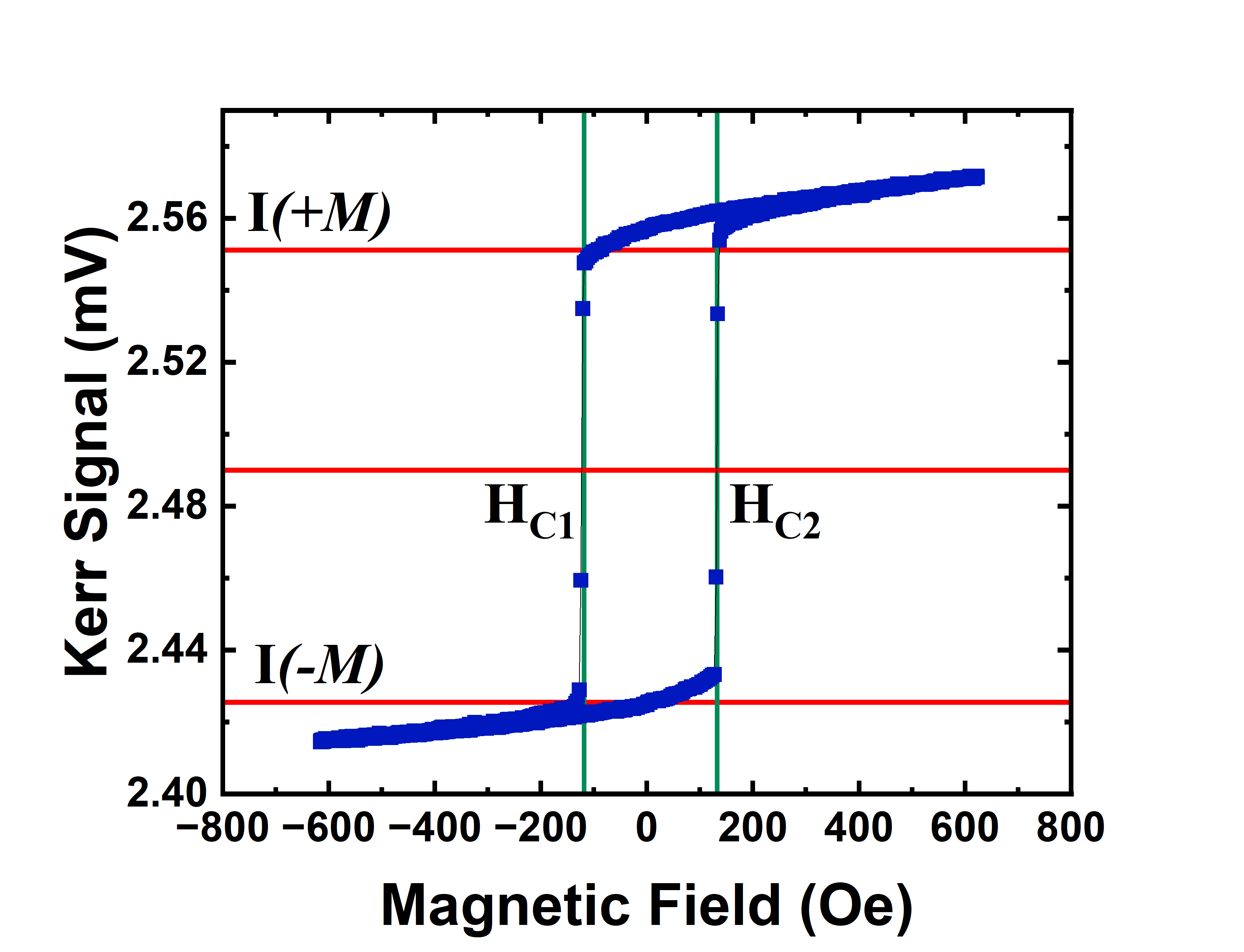}
    \caption{A MOKE hysteresis loop obtained near the cross-polarized condition. The solid red lines correspond to the light intensity when the sample is magnetized along the positive ($+M$) and negative ($-M$) field directions. $H_{C1}$ and $H_{C2}$ denote the coercive fields in the positive and negative magnetic field directions, respectively. Only the hysteretic part of the loop is considered to define the saturation intensities.}
    \label{fig:hys}
\end{figure}
Figure \ref{fig:hys}  shows a typical MOKE hysteresis loop near null-detection obtained using the lock-in techniques. For each hysteresis loop, two measurements are made from the intensities observed at magnetic saturation. First is the MOKE signal defined as

\begin{equation}
    \Delta I = I(+M) - I(-M)
    \label{Kerr signal}
\end{equation}
and the average (zero-field) intensity can also be obtained as 
\begin{equation}
   I = \frac{I(+M) + I(-M)}{2}
   \label{average signal}
\end{equation}

\vspace*{-6mm}

In our experiments, we measure one of the two orthogonal light components using the Wollaston prism. 
For our data analysis, we have considered only the hysteretic portion of the loop as the true Kerr signal and ignored the linear, non-hysteretic portion. Optical misalignment is eliminated as much as possible to maximize the average signal. Hysteresis loop data ($\pm$ 800 Oe) are taken every 10° of the polarizer angle, and the average and MOKE intensities are extracted for each loop. The data is taken for the two polarization modes of incident light and the detector (four combinations in all) and are presented below. The polarization-angle-dependent data is then fitted to the Jones Matrix analysis equations (Table: \ref{Table:alpha_angles}) to obtain the Kerr rotation for both polarization states.

\begin{figure}[t]
    \centering
    \includegraphics[width=0.8\linewidth]{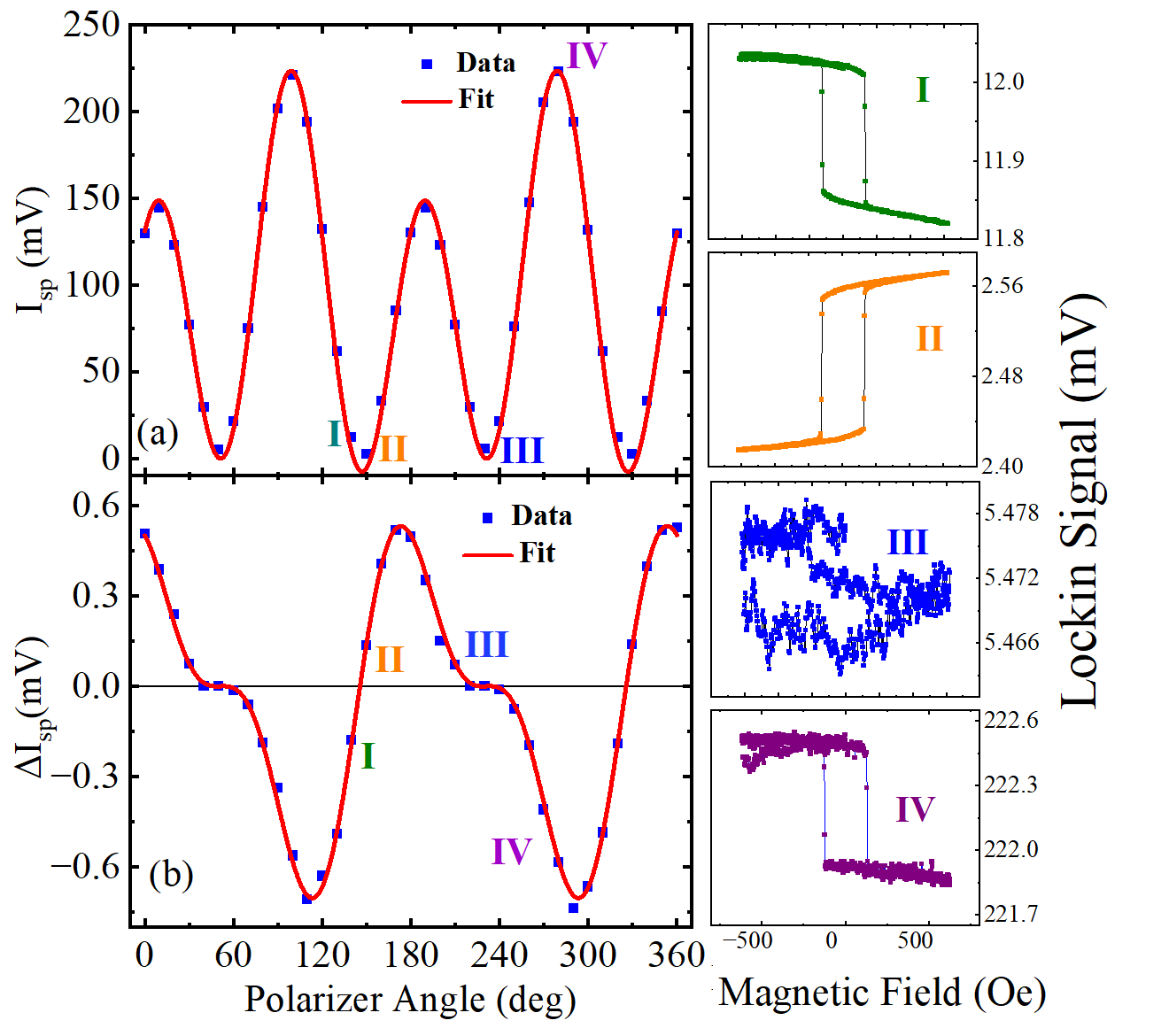}
    \caption{(a) $s$-polarized intensity ($I_s$) and (b) the MOKE signal ($\Delta I_p$) for $p$-polarized incident light. The colored labels indicate the sample orientations corresponding to the hysteresis loops shown in panels (I--IV).}
    \label{fig:p-s}
\end{figure}

\subsection{ \emph{s-p} configuration}

We shall first describe the measurements with the detector set to measure the $s$-polarized intensity for $p$-polarized laser.
In Fig. \ref{fig:p-s}(a)-(b), we show the average intensity ($I_{sp}$) and the MOKE signal ($\Delta I_{sp}$) as the polarization angle is varied a full 360$^\circ$, and then fit to eqns \ref{final E_fv 0} and \ref{eqn:del_Isp}. The equations are reproduced here for clarity, with the polarization angle, $\theta$, measured with respect to the plane of incidence.
\begin{align}
\frac{I_{sp}}{I_0} &= \frac{1}{4}r_s^2 \sin^2(2\theta+\delta_1)
+ r_{sp}^2 \cos^4(\theta+\delta_1)
+ r_s^2 \theta_k^s \cos^2(\theta+\delta_1)\sin(2\theta+\delta_1), \\
\frac{\Delta I_{sp}}{I_0} &=
2r_p^2 \theta_k^p \sin(2\theta+\delta_2)\cos^2(\theta+\delta_3).
\label{eqn:intensity-moke-eqns-with-phases}
\end{align}

$I_0$ represents the laser intensity, which is treated as an unknown value and is not important in our overall analysis. To fit our experimental results, an angular phase was added to every geometric term to account for the offset between the rotation mount reading and the polarizer axis, and to account for the asymmetry to the MOKE data. The polarizer angle shown in Fig. \ref{fig:p-s} and the rest is the nominal rotation mount reading. A phase of $\delta_1$= 35-40$^\circ$  provided  a $R^2$ value, often a metric to describe the goodness-of-fit, of more than 99.6\% for the intensity data (Fig. \ref{fig:p-s}(a)), while two phases ($\delta_2=2\times 34.5 = 69$, $\delta_3$ = 41.4) gave an $R^2=$99.7\% for the MOKE signal. 

 In Table \ref{table:alpha_angles-fit}, we have summarized the equations along with the unnormalized fit values obtained for the various configurations. As the fit values are unnormalized, only the relative fit values have physical significance. For the $s-p$ configuration, the leading term in intensity is the coefficient of the $\sin^2 2\theta$ containing the $r_s^2$ term, whereas the coefficient of the $\cos^4\theta$, $r_{sp}^2$, is the smallest. The ratio of $r_{sp}/r_s$ is 0.0036 which is in the range reported for magnetic materials.  The coefficient of the $cos^2\theta sin2\theta$ term contains the Kerr rotation. 

 In principle, we can obtain the Kerr rotation from the intensity fit alone by comparing the Kerr term to the leading term, but it leads to an abnormally large Kerr rotation of 0.11 radians. This value is orders of magnitude higher than the expected values for Cobalt (0.5-0.6 $milli$radian) as shown in Table \ref{tab:kerr_estimates}. 

 The MOKE signal curve provides a far more accurate and consistent method of determining the Kerr rotation.  Using the $r_s^2$ coefficient value for the average intensity for the unnormalized $r_s^2$, and plugging it into the MOKE signal, we obtained a Kerr rotation of 0.64 milliradians, which is in very good agreement with our $\theta_k^s$ estimate of 0.59 milliradians for s-polarized laser. What we shall demonstrate below is the consistency and accuracy of this method in all measurements. \par

Select hysteresis loops (marked I through IV) are also shown in Fig. \ref{fig:p-s}. First, we observe that the high and low states interchange with respect to the field direction going from I to II. This corresponds to a sign change in the MOKE signal and is due to the $sin 2\theta$ term in the MOKE function.
Also noteworthy is that the signal-to-noise ratio (SNR) is very high under these conditions. This is expected when a null condition (nearly zero average intensity) coincides with the highest MOKE sensitivity, which is determined by the slope in the MOKE signal. This is observed between I and II and, later, 180$^\circ$ away. \par
We also note that loop III has almost no discernible hysteresis signal even though the intensity is close to the null condition. This is, again, due to the angular term in the MOKE signal, which shows a zero value for those specific polarization values and repeats every 180$^\circ$.\par
Loop IV highlights that the higher MOKE signal does not necessarily correspond to the best signal-to-noise ratio (SNR), as the MOKE sensitivity is poor and is close to a minimum. It is generally true that the highest total intensity gives a poor SNR MOKE.

\subsection{\emph{p-p} configuration}

\begin{figure}[H]
    \centering
    \includegraphics[width=0.55\linewidth]{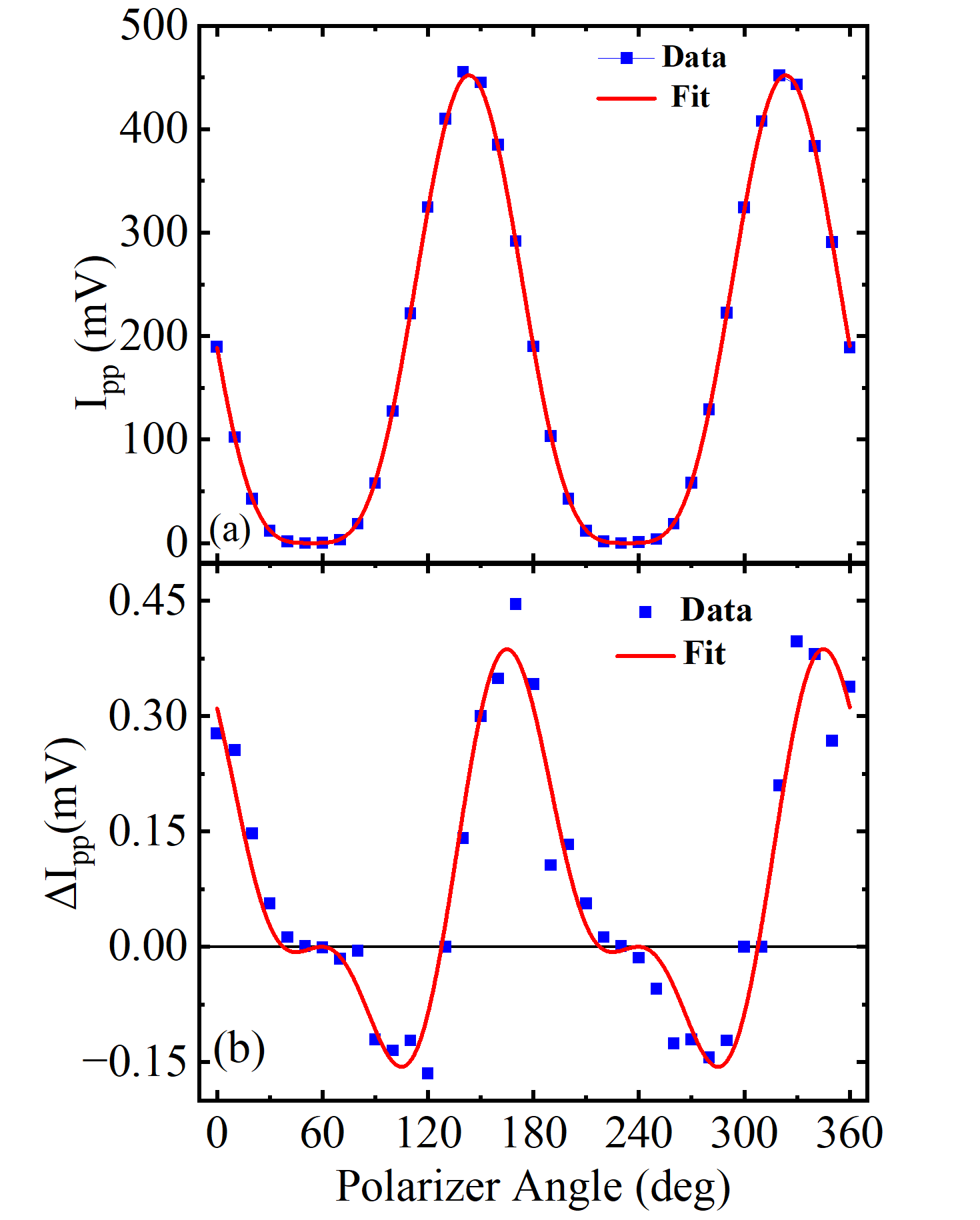}
    \caption{ (a)~$p$- intensity ($I_{pp}$) and (b) the MOKE signal ($\Delta I_{pp}$) for $p$-polarized incident laser.The blue squares represent the experimental data, while the red curve represents the fit to the data}
    \label{fig:p horizontal I_H delIH}
\end{figure}

\vspace*{-6mm}

We next consider the case where the detector measures the $p$-component for a $p$-polarized incident light. In the $p-p$ case, the intensity function is significantly different from the $p-s$ case,
with $cos^4\theta$ being the dominant term.
The MOKE term has the same $sin2\theta cos^2\theta$ dependence as the $s-p$ case, but with an $r_p^2 \theta_k^p$ coefficient. In Fig \ref{fig:p horizontal I_H delIH} we 
present the intensity and MOKE data along with their fit according to the equations above. As with the $s-p$ case, two phase shifts were necessary in the MOKE signal to account for the asymmetry. The measurements were repeated multiple times to verify the asymmetry in the data. Combining the fit parameters from the intensity and MOKE data as shown in Table \ref{table:alpha_angles-fit}, we obtained a Kerr rotation $\theta_k^p$ of 0.46 milliradians, which is again agrees very well with our estimated value of 0.49 milliradians in Table \ref{tab:kerr_estimates}.
Combining the coefficients of the leading term in the intensity equations from the $p-s$ and $p-p$ cases, we can also obtain the ratio of Fresnel coefficients, $|r_p/r_s|=0.796$, which is very close to the available literature value (0.763 from \cite{johnson_optical_1974} for 633 nm).

\subsection{\emph{s-s} and \emph{s-p} configuration}

\begin{figure}[h]
     \centering
     \begin{subfigure}[b]{0.48\textwidth}
         \centering
         \includegraphics[width=\textwidth]{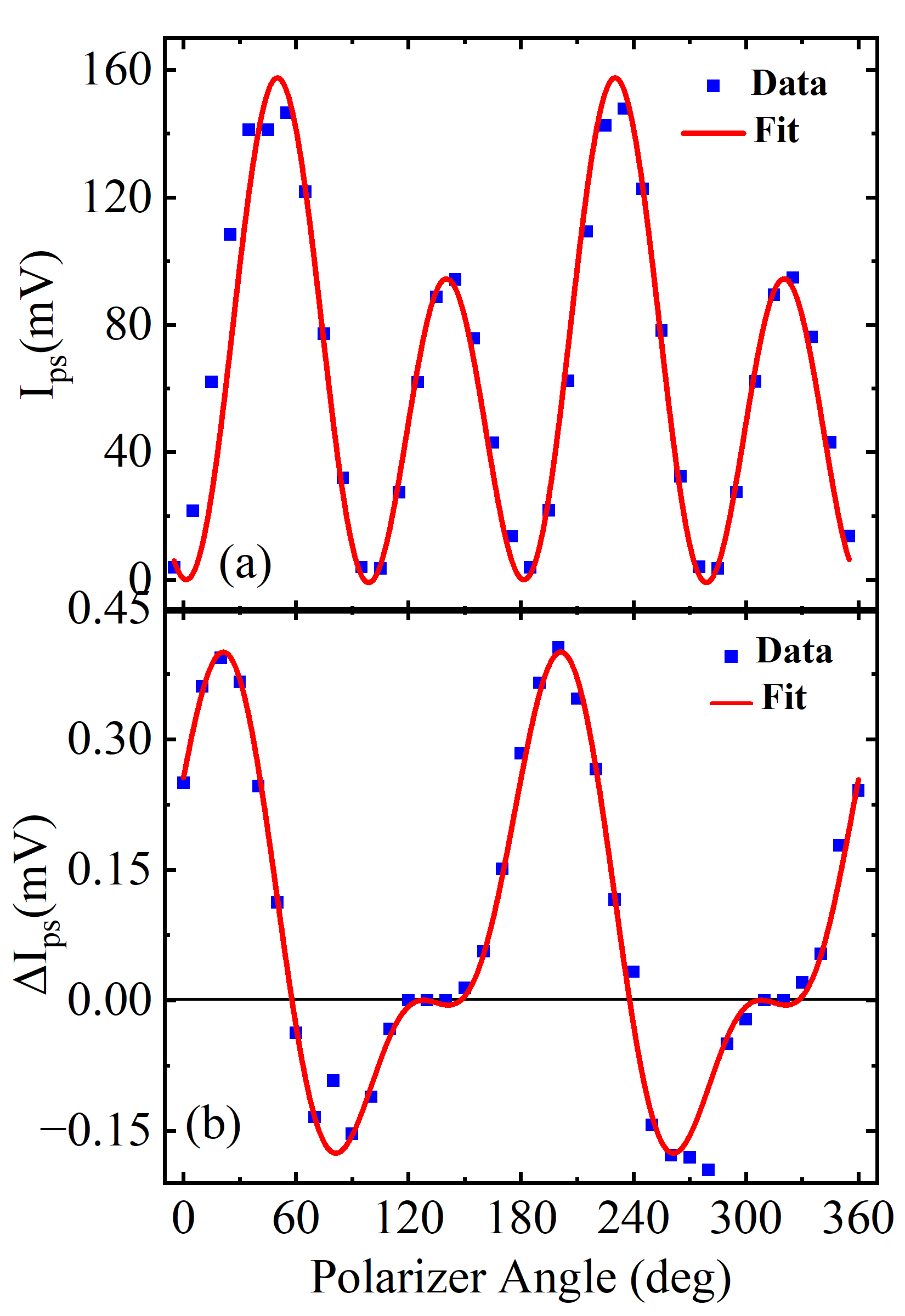}
         \label{fig:y equals x}
     \end{subfigure}
     \begin{subfigure}[b]{0.48\textwidth}
         \centering
         \includegraphics[width=\textwidth]{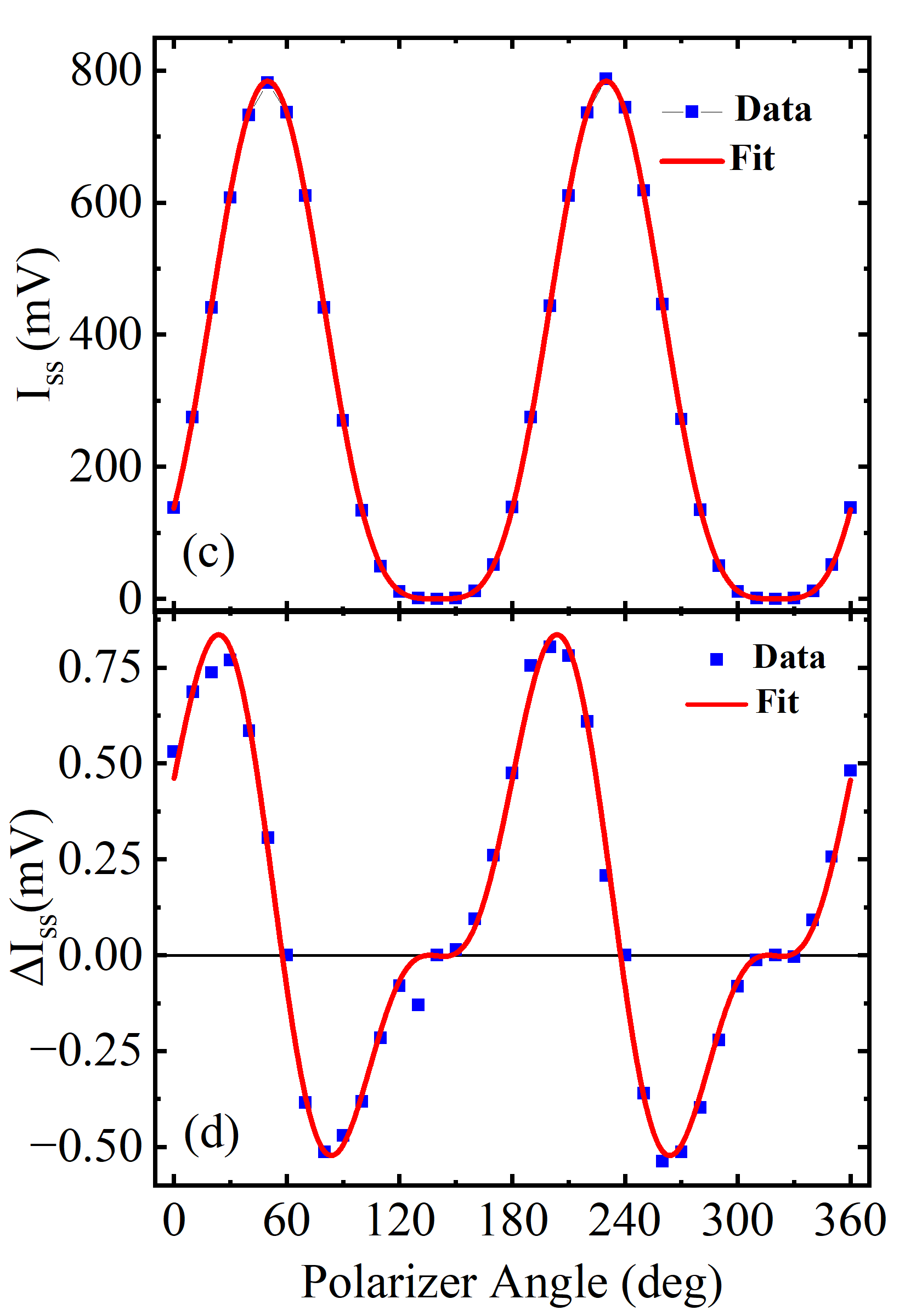}
         \label{fig:y equal x}
     \end{subfigure}
     \caption{ $p$-intensity ($I_{ps}$) and the corresponding MOKE signal ($\Delta I_{ps}$) are shown in panel (a,b), and the $s$ intensity ($I_{ss}$) and MOKE signal ($\Delta I_{ss}$) respectively (c,d) for $s$-polarized laser. The blue points represent the experimental data, while the red curve is the fit to the data}
     \label{fig:s-pol intensity}
     \end{figure}

We can now extend our discussion to the case with $s$-polarized incident laser light. Both detectors settings are shown together in Fig \ref{fig:s-pol intensity}.
Excellent fits are obtained using the equations shown in Table \ref{Table:alpha_angles} by adding phases similar to $p$-polarized light as in eqn. \ref{eqn:intensity-moke-eqns-with-phases}. The Kerr rotation from the two combinations are estimated to be $\theta_k^p$=0.45 mrad by combining fit results from panels (a)-(b), the $p$-detector, and and $\theta_k^s$=0.66 mrad using fits from panels (c)-(d), the $s$-detector. Both values are again in excellent agreement with our estimates (Table \ref{tab:kerr_estimates}). Also, combining the leading terms for the intensity fits (a and c) and we can obtain the Fresnel coefficient ratios, $|r_p/r_s|$ to be 0.78.
\par

\subsection{Summary of all results}

\vspace*{-10mm}

\begin{table}[H]
\centering
\vskip 30truept
\renewcommand{\arraystretch}{1.4}
\setlength{\tabcolsep}{6pt}

\begin{tabular}{|c|c|c|c|c|}
\hline
 & \textbf{\textit{$\bm{p}$-laser}} & fit values & 
\textbf{\textit{$\bm{s}$-laser}} & fit values \\
\hline

$\textbf{\textit{$\bm{p}$-intensity}}$ &
\parbox{3.5cm}{
\[
r_p^2\cos^4\theta
+\frac{1}{4}r_{ps}^2\sin^2(2\theta)
\]
\[
+r_p^2\theta_k^p\sin(2\theta)\cos^2\theta
\]
}
&
\parbox{2.5cm}{
$r_p^2=453-452$

$r_{ps}^2=0-10$

$r_p^2\theta_k^p=-0.22$
}
&
\parbox{3.5cm}{
\[
\frac{1}{4}r_p^2\sin^2(2\theta)
+r_{ps}^2\sin^4\theta
\]
\[
+r_p^2\theta_k^p\sin(2\theta)\sin^2\theta
\]
}
&
\parbox{2.5cm}{
$r_p^2=501-489$

$r_{ps}^2=0-7$

$r_p^2\theta_k^p=61-63$
}
\\
\hline

$\textbf{\textit{$\bm{p}$-MOKE}}$ &
\parbox{3.5cm}{
\[
2r_p^2\theta_k^p
\sin(2\theta)\cos^2\theta
\]
}
&
\parbox{2.5cm}{
$r_p^2\theta_k^p=0.21$

$\boxed{\theta_k^p=0.00046}$
}
&
\parbox{3.5cm}{
\[
2r_p^2\theta_k^s
\sin(2\theta)\sin^2\theta
\]
}
&
\parbox{2.5cm}{
$r_p^2\theta_k^p=0.22$

$\boxed{\theta_k^p=0.00045}$
}
\\
\hline

$\textbf{\textit{$\bm{s}$-intensity}}$ &
\parbox{3.5cm}{
\[
\frac{1}{4}r_s^2\sin^2(2\theta)
+r_{sp}^2\cos^4\theta
\]
\[
+r_s^2\theta_k^s
\sin(2\theta)\cos^2\theta
\]
}
&
\parbox{2.5cm}{
$r_s^2=736.3$

$r_{sp}^2=0.01$

$r_s^2\theta_k^s=-75.1$
}
&
\parbox{3.5cm}{
\[
r_s^2\sin^4\theta
+\frac{1}{4}r_{sp}^2\sin^2(2\theta)
\]
\[
+r_s^2\theta_k^s
\sin(2\theta)\sin^2\theta
\]
}
&
\parbox{2.5cm}{
$r_s^2=785-783$

$r_{sp}^2=0-12.9$

$r_s^2\theta_k^s=0-32.5$
}
\\
\hline

$\textbf{\textit{$\bm{p}$-MOKE}}$ &
\parbox{3.5cm}{
\[
2r_s^2\theta_k^s
\sin(2\theta)\cos^2\theta
\]
}
&
\parbox{2.5cm}{
$r_s^2\theta_k^s=0.47$

$\boxed{\theta_k=0.00064}$
}
&
\parbox{3.5cm}{
\[
2r_s^2\theta_k^s
\sin(2\theta)\sin^2\theta
\]
}
&
\parbox{2.5cm}{
$r_s^2\theta_k=0.52$

$\boxed{\theta_k=0.00066}$
}
\\
\hline

\end{tabular}

\caption{Intensity and MOKE signal equations and the unnormalized experimental fit values for the different cases examined in this work.}
\label{table:alpha_angles-fit}
\vskip 12truept
\end{table}

In Table~\ref{table:alpha_angles-fit}, we show the results from the unnormalized fit parameters along with the equations shown earlier in Table \ref{Table:alpha_angles} for the different cases. Apart from the excellent agreement in the Kerr rotation between our theory and calculations mentioned above, the most noteworthy observation is that several of the fit parameters are reported with a range of values with acceptable fit. For example, we obtain a good fit by ignoring the $r_{ps}^2$ (or $r_{sp}^2$)  term completely in all cases. Ignoring the $r_{ps}^2$ terms results in adjusted values for the other terms. For example, if we restrict $r_{sp}^2=0$ in the $p-s$ setting, the leading $r_p^2$ converges to an unnormalized fit value of 501 with a 94\% $R^2$ goodness of fit. Keeping $r_{sp}^2$ as a free parameter gives $r_{sp}^2=7$ and $r_p^2=489$ and improves the $R^2$ value to 94.6\%. As is obvious, the improvement in the fit is marginal, and all values within the range is, therefore, acceptable. Such is observed consistently in the other cases as well.

\section{CONCLUSIONS}

We developed a variable-angle polarizer method longitudinal MOKE setup to estimate the Kerr rotation accurately by fitting the intensity and MOKE data as a function of the polarizer angle to equations obtained via a Jones matrix analysis. Using a cobalt thin film as a sample, we analyzed the data for different laser and detector configurations to obtain the $s$ and $p$ Kerr rotation. The estimated values for the Kerr rotation are $0.46~\mathrm{mrad}$ for the $p$-polarization and $0.66~\mathrm{mrad}$ for the $s$-polarization. Such values are in excellent agreement with the expected values that demonstrates the accuracy of the method.

\section{ACKNOWLEDGEMENTS}
This work was supported by the NSF CAREER grant (ECCS 1846829)

\newpage
   
\nocite{*}
\bibliography{aiptemplate}

\end{document}